\documentstyle[11pt,newpasp,twoside,epsf]{article}
\def\spose#1{\hbox to 0pt{#1\hss}}
\def\lta{\mathrel{\spose{\lower 3pt\hbox{$\mathchar"218$}}
     \raise 2.0pt\hbox{$\mathchar"13C$}}}
\def\gta{\mathrel{\spose{\lower 3pt\hbox{$\mathchar"218$}}
     \raise 2.0pt\hbox{$\mathchar"13E$}}}
\def\etal{{\it et al.\ }}

\def\edcomment#1{\iffalse\marginpar{\raggedright\sl#1\/}\else\relax\fi}
\reversemarginpar

\begin{document}
\title{Modelling Tools: Population and Evolutionary Synthesis}
 \author{Uta Fritze -- v. Alvensleben}
\affil{Universit\"atssternwarte G\"ottingen, Geismarlandstr. 11, D -- 37083 G\"ottingen, Germany}

\begin{abstract}
I review the basic concepts for the spectrophotometric and chemical 
evolution of galaxies, contrast various approaches and discuss their 
respective advantages and shortcomings, both for the interpretation 
of nearby and high redshift galaxies. Focus is on recent attempts to 
include gas and dust into galaxy evolution models and to account 
for the links among stars, gas and dust. Chemically consistent models 
are described that try to cope with extended stellar 
metallicity distributions observed in local galaxies and with subsolar 
abundances in young galaxies. 
\end{abstract}


\section{Introduction}
For local group galaxies that can be resolved into individual stars with HST, we have seen impressively, e.g. in Carme Gallart's contribution, how the comparison of deep color magnitude diagrams with isochrones calculated from stellar evolutionary tracks allows to trace back their {\large\bf S}tar {\large\bf F}ormation {\large\bf H}istory ({\bf SFH}) over the last 1 -- 3 or even more Gyr. Accessing earlier phases becomes increasingly difficult since stars on the Main Sequence are at or beyond the faint limits of these observations. Galaxies beyond the local group remain inaccessible for those techniques even with 10 m telescopes. 

In this contribution, we now turn to unresolved galaxies and to attempts to derive information from their integrated light about their present star, gas, and dust content as well as about their evolutionary path to this present state. 

Modelling techniques are presented in Sect. 2 for the description of the stellar light including absorption and emission effects of gas and dust. In Sect. 3, I present first attempts to consistently account for the physical links between those 3 components in galaxy models and I present a brief selection of recent results in Sect. 4. Sect. 5 concludes with a brief outlook.

\section{Modelling Techniques}
\subsection{Basic Principles}
Two fundamentally different approaches have been developed for the interpretation of integrated spectrophotometric properties of galaxies: {\large \bf P}opulation {\large \bf S}ynthesis ({\bf PS}) and {\large \bf E}volutionary {\large \bf S}ynthesis ({\bf ES}). Both of them need a complete input library in terms of (observed or model atmosphere) stellar spectra, star cluster or isochrone spectra. A long wavelength basis is required if information about the SFH over cosmological timescales is to be obtained. 

While in the PS approach, a minimisation algorithm selects a linear combination of input spectra that best fits the observed spectrum of a local galaxy or a galaxy at known redshift z. Codes may give particular weight to wavelength regions more sensitive than others to age or metallicity and include physical boundary conditions like some IMF or reasonable number ratios of stars in different evolutionary stages (cf. Whipple 1935, Alloin \etal 1971ff, Faber 1972ff, O'Connell 1976ff, Pickles 1985ff, Bica \& Alloin 1986ff, ...). Advantages of PS are that very good spectral fits are possible and unexpected or discontinuous SFHs can be discovered. The basic shortcoming of this approach is that it only provides a {\sl status quo} description without information about the evolutionary path. There is still debate about the uniqueness of PS solutions. 

ES starts from a gas cloud of given initial mass and chemical composition, specifies an IMF and a SFH to successively build up the stellar population of a model galaxy. The code basically is a book-keeping algorithm that -- on the basis of a complete set of stellar evolutionary tracks -- follows the evolution of the stellar population through the HR diagram. 
Assigning photometric properties like luminosities, colors, and absorption features, or spectra to every point or cell in the HR diagram allows to follow the photometric or spectral evolution, respectively, of the model galaxy from the onset of SF over a Hubble time (cf. Tinsley 1972ff, Searle, Sargent \& Bagnuolo 1973, Huchra 1977, Bruzual 1982ff, Arimoto \& Yoshii 1986ff, Rocca-Volmerange \& Guiderdoni 1987ff, Buzzoni 1989, F.-v.A. 1989ff, Charlot \& Bruzual 1991ff, Worthey 1994, Poggianti 1997, Leitherer \etal 1999). 
Advantages of ES are that it directly desribes the time evolution to and after the present state. Evolutionary consistency is guaranteed and models feature a large analytical potential. Luminosity contributions at any wavelength from stars of different masses, spectral types, luminosity classes, and evolutionary phases are given in their time evolution. The ES approach allows for combination with a cosmological model characterised by its parameters (${\rm H_o,~\Omega_o,~\Lambda_o}$) and an assumed redshift of galaxy formation ${\rm z_f}$. Major shortcomings of ES are the facts that good fits to observed galaxy spectra are more difficult to obtain than with PS and that unexpected SFHs are difficult to discover. This points out the complementary characters of the ES and PS approaches. While for normal undisturbed galaxies average type specific SFHs are known, for disturbed or peculiar galaxies PS may provide a first guess for their stellar content. To explore their evolutionary past and future and to allow for cosmological considerations, the PS solution provides a useful guide for follow-up ES modelling. For a more extensive review on synthesis techniques see F.-v.A. 1994. 

In 1996, a workshop devoted to the comparison of various ES and PS codes showed broad agreement in general and a number of discrepancies in detail. The latter turned out to be largely due to differences in the input physics used. Of prime importance is the completeness of the stellar tracks up to the latest evolutionary phases, e.g. the thermal pulsing AGB phase (cf. Leitherer \etal 1996). 

\subsection{Inclusion of Gas}
Both PS and ES models, as far as presented above, describe the integrated light of the {\bf stellar} component of galaxies, i.e. they yield absorption line spectra. Emission lines observed in the spectra of SFing galaxies are evidence for the sometimes very strong contribution of an ionised gas component (HII regions). 

\medskip\noindent
{\bf 2.2.1. Gaseous emission} \\
Gaseous emission in terms of lines and continuum is readily included into ES and PS models on the basis of the number of Lyman continuum photons emitted by all the hot O- and B-stars present in a model galaxy at a certain time (e.g. Huchra 1977, Mas-Hesse \& Kunth 1991, Rocca-Volmerange \& Guiderdoni 1988, Kr\"uger \etal 1991, 1994, 1995, Leitherer \& Heckman 1995). Ionisation-recombination equilibrium in ionisation-bounded Str\"omgren spheres that are optically thick in the Lyman lines is assumed as well as values for the electron temperature and density. Then the continuous emission from free-free transitions and recombination of H$^+$, He$^+$, He$^{++}$ is obtained from atomic physics are the fluxes in the hydrogen lines. Other important lines are calculated from the flux in H$_{\beta}$ using line ratios either from photoionisation models or from observations. These metal-to-hydrogen-line flux ratios, of course, severely depend on the metallicity of the gas. 
A recent redetermination of the number of Lyman continuum photons emitted by hot stars including line-blanketing, non-LTE and wind effects is given by Schaerer \& deKoter 1997. 

\medskip\noindent
{\bf 2.2.2. Effects on the broad band photometry} \\
In models for galaxies with a starburst superimposed on an undisturbed galaxy's normal SFR, gaseous emission is seen to significantly affect broad band colors. As shown in Kr\"uger \etal 1995, in bursts increasing the stellar mass of a galaxy by 10 \%, gaseous line emission accounts for $\geq 50$ \% of the broad band fluxes in the optical (UBVRI), while gaseous continuum emission makes up $\sim 20$ \% of the NIR (JHK) fluxes. 
Moreover, starburst galaxy models show that there is a time delay of order 40 Myr until the maximum flux in the gaseous emission is reached. For dwarf galaxies with their typical burst durations of order $10^6$ yr the SFR has already declined to $\sim 20$ \% of its original value by the time until the maximum H$_\alpha$ luminosity is reached. Implications for a determination of the SFR from ${\rm L_{H_{\alpha}}}$ in these cases are profound. 

\medskip\noindent
{\bf 2.2.3. Gaseous absorption} \\
Within a gas rich SFing galaxy, the flux of the stellar component is largely selfabsorbed below the Lyman limit at 912 \AA \ by intrinsic HI, producing a pronounced break in the spectrum that for galaxies at redshifts z $> 2$ is redshifted into the optical. Intergalactic HI, mostly in the form of Ly$\alpha$ clouds stochastically distributed along sightlines to distant galaxies, provides additional absorption of the emitted galaxy light below rest-frame Ly$\alpha$ at 1216 \AA. The longer the sightline to a galaxy, the stronger becomes this attenuation (Madau \etal 1996). For galaxies at z $\gta 2.5, ~3.5, ~4.5, ~.~.~.~$, the Lyman Break is shifted longwards of the U, B, V, ... filters, respectively. This causes galaxies to drop out of deep short wavelength exposures while visible in longer wavelength bands. It hence provides a powerful tool to select promising galaxy candidates at high and very high redshifts from deep multiband imaging (cf. Steidel \etal 1995). Spectroscopy of an appreciable number of those Lyman Break galaxies has confirmed their redshift estimates. 

\subsection{Inclusion of Dust}
Now that we have seen the effects of gas absorption and emission on the integrated galaxy light, we turn to the effects of dust which, as well, come both in absorption and emission. For many years, galaxy modelers used to ask observers to deredden their galaxy spectra with respect to both foreground and internal extinction. Comprehensive galaxy models by today, however, attempt to include dust and study its effects over a long wavelength range. 

\medskip\noindent
{\bf 2.3.1. Dust absorption} \\
With a given dust content, expressed e.g. in terms of a value for ${\rm E_{B-V}}$ and an assumped extinction law, the effect of dust absorption is accounted for in synthetic galaxy spectra via ${\rm F_{gal}^{dust} (\lambda) = F_{gal} (\lambda) \cdot 10^{-0.4 \cdot A_{\lambda}}}$ with ${\rm A_V = {\cal R} \cdot E_{B-V}}$ and ${\rm A_{\lambda} / A_V}$ given by the extinction/absorption law. 
Many details and a wealth of interesting physics are ignored in this zero-order approach, as e.g. the geometry of the distributions of old and young stars and dust, the physical relation between gas content, metallicity and dust, the temperature, chemical composition, grain sizes of the dust, etc. 
Tying the amount of extinction to the evolving gas and metal content, Fioc \& Rocca -- Volmerange 1997 and M\"oller \etal 1999 present ES models with dust absorption for various galaxy types. Their present-day values for ${\rm E_{B-V}}$ are in agreement with observed local averages for the different galaxy types (Goudfrooij \etal 1994, Ferrari \etal 1999 for E/S0s, Gonzalez \etal 1998 for spirals). Individual galaxies show considerable scatter around these average values. Only on a long wavelength basis may a comparison between models and observations provide conclusive information about the amount of dust absorption. Extrapolation of this approach into the Early Universe with its intense radiation field remains uncertain. 

\medskip\noindent
{\bf 2.3.2. Dust emission} \\
Energy conservation requires the total amount of flux absorbed in the UV and optical to be reemitted at long wavelengths (FIR and submm) 
\begin{center} ${\rm L_{FIR-submm} \sim \int^{2 \mu}_{90 \AA} (F_{gal}(\lambda) - F_{gal}^{dust}(\lambda)) \cdot d{\lambda}}$\end{center}

Thermal emission of dust, eventually heated by young stars or shocks, produces a blackbody spectrum for the appropriate temperature(s) of the dust (phases) plus characteristic line and (PAH-)band contributions (cf. Sturm \etal 2000). Models including dust emission are e.g. presented by Franceschini \etal 1994, Silva \etal 1998, Devriendt \etal 1999.

\section{Links between Stars, Gas, and Dust}
\subsection{Chemically consistent models}
As first described quantitatively by B. Tinsley in the 70s, the gas in galaxies is continuously enriched by metals produced in stars and set free towards the end of their lifes by stellar winds, planetary nebulae or supernovae. Hence, in galaxies that always have SF extending over timescales that are long compared to stellar lifetimes, successive generations of stars will be born with increasing initial metallicities. This is observed in the extended metallicity distributions of stars in elliptical galaxies and bulges -- covering more than a factor 10 in metallicity Z with ${\rm \langle Z \rangle \sim \frac{1}{2} \cdot Z_{\odot}}$ (McWilliam \& Rich 1994, Sadler \etal 1996, Ramirez \etal 2000) -- and of G-, K-, and M-dwarf stars in the solar neighbourhood -- also covering a factor 10 in Z with ${\rm \langle Z \rangle \sim 0.7 \cdot Z_{\odot}}$ (Rocha -- Pinto \& Maciel 1996). 

In principle, PS using a star cluster library covering the full range of ages and metallicities and extending over a long enough wavelength range should be able to recover the intrinsic metallicity distribution of the stars in its best fit solution to an integrated galaxy spectrum (cf. Bica \etal 1988). \\
With the availability of complete sets of input physics for various metallicities, we developed {\large \bf C}hemically {\large \bf C}onsistent {\large \bf E}volutionary {\large \bf S}ynthesis {\bf (CCES)} models in G\"ottingen. They follow the enrichment of the gas and account for the increasing initial metallicity of successive generations of stars by using several sets of input physics -- stellar evolutionary tracks, stellar yields, model atmospheres, color and index calibrations -- for metallicities in the range ${\rm 0.0001 \leq Z \-\leq \- 0.05}$ and yield stellar metallicity distributions for elliptical and Sb galaxies in agreement with observations. \\
For the {\bf CCES} models average galaxy wide SFHs  are determined by the requirement that after $\sim 12$ Gyr of evolution they agree with observations of nearby galaxies in terms of colors, luminosities, absorption features (for E/S0s), emission line strengths (spirals), and template spectra from UV to NIR. In order to reach the observed galaxy colors and luminosities despite the contributions from intrinsically bluer and more luminous lower metallicity subpopulations, CCES models have to use SFHs somewhat different from those in models using solar metallicity input physics only. This causes a difference in the evolution of galaxies that increases with lookback time, even for those types which by today reach close to solar average abundances. 

The well-known {\bf age -- metallicity degeneracy} (Worthey 1994) describes the fact that two single age single metallicity stellar populations, one of which is younger by a factor 2 and more metal rich by a factor 3 than the other, are indistinguishable in terms of optical broad band colors and spectral continuum shape. To resolve this ambiguity less age dependent NIR colors or absorption features specifically sensitive to age (like ${\rm H_{\beta}}$) or metallicity (like Fe5270 or Fe5335) have to be considered. 
CCES models show that in normal Hubble types of galaxies which, in this respect, are well approximated by closed box models, {\bf age and metallicity of the stellar population are coupled} and {\bf the coupling depends on the SFH} (cf. F.-v.A. 2000).

\section{Results from CCES models}
\subsection{Chemically Consistent Spectrophotometric Evolution}
The fact that stellar subpopulations of different ages have different metallicities and dominate different regions of the spectrum causes galaxies to feature different metallicities at different wavelengths. In M\"oller \etal 1997 we present a quantitative analysis of this wavelength dependence of luminosity weighted metallicities in ellipticals and spirals. \\
CCES models can be combined with a cosmological model (e.g. ${\rm H_o=50, \- \Omega_o=1}$)  to study the redshift evolution of galaxy colors and luminosities. Comparison with solar metallicity models shows that cc spiral models become brighter by $\sim 1$ ($>0.5$) mag in B (K) at redshifts ${\rm z \gta 0.5}$. We thus expect higher numbers of early and even late-type spirals to contribute to faint galaxy counts and, hence, less of a Faint Blue Galaxy Excess (cf. F.-v.A. 1999, M\"oller \etal {\sl in prep}). CCES models for spirals show good agreement both in terms of luminosity and color evolution and in SFRs with the first set of spectroscopically confirmed Lyman Break Galaxies from the HDF (Lowenthal \etal 1997) over the redshift range from ${\rm z \sim 2}$ to ${\rm z \sim 3.5}$. On the basis of our 1-zone models, it cannot be decided, however, if this SF occurs in the bulges or all over the disks of those young spirals. 
From the weakness of O-, B-stellar wind lines in KECK HIRES spectra Trager \etal 1997 derive metallicities for Lyman Break Galaxies at ${\rm z \sim 3 - 4}$ around ${\rm \frac{1}{10} \cdot Z_{\odot}}$ in agreement with the predictions of our CCES models for bright early type spirals. 

\subsection{Chemically Consistent Chemical Evolution}
Stellar yields for individual chemical elements ${\rm X_i}$ show a very complex behaviour as a function of metallicity  and stellar mass (cf. Figs 1, 2 in Lindner \etal 1999). SFHs that after a Hubble time gave agreement of our spectrophotometric CCES models with observations immediately produced absolute oxygen abundances in good agreement with observed characteristic HII region abundances (i.e. measured at ${\rm \sim 1~ R_e}$) of the respective spiral types without any additional free parameters or scaling. The redshift evolution of abundances in CCES models differs from that in models using solar metallicity input physics only in a way characteristic for every element. 

{\large \bf D}amped {\large \bf L}y$\alpha$ {\large \bf A}bsorption is caused by intervening high column density gas (log N(HI) ${\rm \geq 20.3~cm^{-2}}$) along our lines of sight to distant quasars. The radiation damped Ly$\alpha$ lines are always associated with a number of low ionisation lines of elements like Al, Si, S, Fe, Ni, Mn, Cr, Zn at the same redshift. High resolution spectra resolving the velocity structure in these lines (typically ${\rm \sim 5 - 10~km~s^{-1}}$) allow for precise abundance determinations. In recent years, those have become available for a large number of absorbers over the redshift range from ${\rm z \sim 0}$ to ${\rm z > 4.4}$, i.e. over more than 90 \% of the age of the Universe (e.g. Pettini \etal 1999). Conventionally, DLA absorption was thought to arise in (proto-)galactic disks (Wolfe 1995). Line profile shapes are consistent with this view and indicate that DLA disks at ${\rm z \sim 2 - 3}$ already have rotation velocities and, hence, masses comparable to those of local spiral disks (Prochaska \& Wolfe 1997ff). Nevertheless, other possible origins for DLA systems are discussed like starbursting dwarf or low surface brightness galaxies at low z (Matteucci \etal 1997, Jimenez \etal 1999) and subgalactic fragments bound to merge at high z (Haehnelt \etal 1998). 

In Lindner \etal 1999, we compare our CCES models for spirals (including SNIa contributions from carbon deflagration in white dwarf binaries) to DLA abundances. The basic result from the comparison of all 8 elements with significant numbers of abundance determinations is that over the full redshift range from ${\rm z \gta 4.4}$ to ${\rm z \sim 0}$ cc spiral models are well consistent with the DLA data. Hence, from the point of view of abundance evolution, DLA absorption may well be caused by normal spiral galaxies, allthough other possible origins are not excluded. DLA abundances at high redshift are in evolutionary consistency with present-day HII region abundances. The decreasing gas content explains the lower rate of incidence of DLA galaxies at low redshift as compared to expectations from high redshift and it causes a change in the DLA absorber population with redshift. While at high z all spiral types from Sa through Sd seem to contribute to the DLA population, the gas poor and high metallicity Sa and Sb galaxies seem to drop out of DLA samples towards lower redshift. This change with redshift that we predict for the DLA absorber population is consistent with and makes us understand the non-detection of many DLA absorbers in deep searches. The luminosities that our CCES models predict (${\rm B \sim 25}$, ${\cal R} \sim 25$, ${\rm K \sim 22}$ mag, cf. F.-v.A. \etal 1999) make DLA galaxies challenging targets for 10 m telescopes. 

The ISM abundance evolution as a function of redshift given by our spiral models can be used to predict metallicities of stars, star clusters, and Tidal Dwarf Galaxies forming in the course of mergers and the starbursts they can trigger (F.-v.A. \& Gerhard 1994a, b). Spectroscopy of those objects, in turn, provides an independent cross check of their parent galaxies' abundance evolution and has confirmed our predictions in several cases (Schweizer \& Seitzer 1998, Duc \& Mirabel 1998). The metallicity of all these objects is crucial for age-dating, mass-to-light ratios and for their future luminosity and color evolution (F.-v.A. \& Burkert 1995, F.-v.A. 1998, 1999). 

\subsection{SSP Models}
Having available the input physics for different metallicities, not only galaxy but also single burst single metallicity models ($=$ {\large\bf S}imple {\large\bf S}tellar {\large\bf P}opulation models) can be calculated (Kurth \etal 1999). Those not only provide the luminosity, color and absorption index evolution directly comparable to star cluster data but also a useful tool for the implementation into any kind of dynamical or cosmological galaxy formation and evolution model if only it contains a SF criterion. Any kind of composite stellar population, no matter how complex its SFH may be, is readily expanded into a series of SSPs and its luminosity evolution is obtained from a superposition of those of the respective SSPs. In Contardo \etal 1998 we present a first attempt to supplement the hydro-cosmological formation and evolution of a protogalaxy with observables using SSP results. We have recently completed a new set of SSP models including the detailed spectral evolution (Schulz \etal, {\sl in prep.}). They are based on Padova isochrones that extend the stellar evolutionary tracks used by Kurth \etal to include the thermal pulsing AGB phase.
 
\section{Outlook}
Contrasting the two basic approaches to model galaxy spectra -- evolutionary and population synthesis -- I pointed out their respective advantages and shortcomings as well as their complementary character when dealing with complex or unexpected star formation histories. Originally, both approaches describe the stellar content. Contributions of gas and dust, both in emission and absorption, are now being included by various groups. Very first attempts are reported to consistently account for the physical links between these 3 galaxy components. They still are restricted to very simplified 1-zone closed box models with only one gas phase to keep the number of parameters restrictable. Observations allowing by today to resolve galaxies out to large distances call for spatially resolved models that include the dynamics of all components and phases and also include effects of environment or even embed a galaxy into its cosmological context. The increasing number of parameters in those kinds of models can only be balanced by comprehensive observations of spectral, chemical and dust properties at all wavelengths. 

%\begin{figure}[t]
%\plottwo{fritze1a.eps}{fritze1b.eps}
%\caption{(a) Age distribution of {\bf all} clusters, (b) LF evolved to 12 Gyr by differential fading for 
%{\bf young} clusters in the Antennae. The vertical arrow indicates the completeness limit.}
%\end{figure}
%\begin{figure}[t]
%\plottwo{fritze2a.eps}{fritze2b.eps}
%\caption{(a) LF of old GCs with Gaussian for GCs in (Milky Way + M31).  
%(b) MF of the young clusters.}
%\end{figure}

\acknowledgements
It's a pleasure to thank the organisers for a very stimulating conference and I gratefully acknowledge partial travel support from ESO.

\end{document}